\documentclass[conference]{IEEEtran}
\IEEEoverridecommandlockouts
\usepackage{config}
    
\begin{document}

\title{A Neurosymbolic Prolog Skill for LLM-Driven Service Placement}

\author{
\IEEEauthorblockN{
Jacopo Massa$^{*}$\thanks{$^{*}$Corresponding author: jacopo.massa@di.unipi.it.},
Giuseppe Bisicchia,
Patrizio Dazzi,
Antonio Brogi
}
\IEEEauthorblockA{
\textit{Department of Computer Science, 
University of Pisa, Pisa, Italy}
}
}

\maketitle

\begin{abstract}
Service placement in the cloud-edge continuum requires assigning application components to heterogeneous resources under multiple constraints, including latency, locality, and policy requirements. Existing approaches rely on optimisation models or heuristics that require explicit modelling, while neural methods lack transparency and formal guarantees. 
This work proposes a neuro-symbolic alternative based on a Prolog \textit{skill}, a reusable interface for schema-constrained fact generation and querying, for constraint-aware placement. The skill enables a language model to structure placement intent into symbolic facts, rules, and queries, while delegating validation and reasoning to Prolog.
This design bridges high-level intent and formal constraint evaluation, enabling inspectable and policy-aware placement decisions in cloud-edge environments.
\end{abstract}

\begin{IEEEkeywords}
cloud-edge continuum, service placement, neuro-symbolic reasoning, declarative programming.
\end{IEEEkeywords}

\section{Introduction}
\label{sec:intro}

Service placement in cloud-edge environments requires assigning application components to heterogeneous resources while satisfying constraints related to latency, locality, resource availability, and policy compliance. The compute continuum extends cloud computing towards fog and edge resources, enabling distributed execution across nodes with different capabilities and operational conditions \cite{shi2016edge,satyanarayanan2017emergence,bisicchia2023customisable,bisicchia2022distributed}. In this context, the combinatorial nature of the problem and the diversity of constraints make placement decisions increasingly complex.

Existing approaches follow three main directions. Optimisation-based methods provide formal guarantees but require explicit modelling and specialised expertise, while heuristic and learning-based approaches improve scalability at the cost of interpretability and robustness \cite{brogi2020fogtorch,mahmud2018fog}. Logic-based systems, such as those based on Prolog, offer declarative modelling and explainability, but require manual construction of knowledge bases.

Recent advances in language models enable a new direction, where natural language descriptions can be translated into structured representations suitable for symbolic reasoning. However, fully neural approaches lack guarantees, while hybrid approaches often keep the solver separate, leaving unresolved the gap between high-level intent and formal modelling.

This work proposes a Prolog-centric perspective, where symbolic reasoning constitutes the core decision mechanism. In particular, the goal is not to develop a new Prolog-based reasoner, but to define a reusable \textit{skill}\footnote{The complete format specification is available at: \url{https://agentskills.io}} that enables language models to interact with symbolic reasoning systems for placement tasks. A language model is used to structure placement intent into a symbolic representation, while Prolog performs validation, constraint reasoning, and explanation. The goal is to bridge the gap between high-level intent and formal reasoning in a transparent and inspectable way.

\noindent This work proposes the following main contributions:

\begin{itemize}
	\item a Prolog-based \textit{skill} to enable language models to structure, validate, and reason over placement constraints,
	\item the integration of symbolic guardrails for validating generated knowledge bases, and
	\item an approach to explainable service placement based on constraint reasoning.
\end{itemize}
\section{Background and Related Work}
\label{sec:related}

\subsection{Service Placement}

Service placement in the cloud-edge continuum is typically formulated as a constrained optimisation problem, mapping application components to heterogeneous resources under capacity, latency, and policy constraints~\cite{mahmud2018fog,bisicchia2023customisable}. Existing approaches rely on optimisation, heuristics, or graph-based strategies tailored to specific goals and assumptions.

Optimisation-based formulations, such as Mixed-Integer Linear Programming, provide formal guarantees but require explicit modelling and are difficult to adapt to dynamic or intent-driven scenarios. Learning-based approaches, including reinforcement learning and graph-based scheduling, improve scalability but often lack transparency and robustness.

Recent work has explored more advanced formulations, such as graph partitioning and availability-aware placement that explicitly model service dependencies and infrastructure structure~\cite{lera2024availability}. While improving QoS and scalability, these approaches still depend on explicit modelling and do not support high-level intent specification.

Declarative approaches address this limitation by modelling placement through logic. Prolog-based frameworks such as FogTorch enable constraint-based reasoning for IoT and fog deployments~\cite{brogi2020fogtorch}, later extended to continuous reasoning in distributed environments~\cite{forti2022declarative,bisicchia2024continuous}. A data-aware declarative formulation of placement in the cloud-IoT continuum has been proposed in prior work, encoding infrastructure, application, and data constraints as logical facts and rules~\cite{massa2022dataaware}. More recent approaches combine declarative reasoning with optimisation techniques to improve scalability~\cite{massa2025edgewise}.

These approaches show the effectiveness of Prolog in modelling placement constraints and enabling explainable reasoning, but still rely on manually defined knowledge bases and do not translate high-level intent into formal models.

\subsection{Neuro-Symbolic Reasoning}

Neuro-symbolic AI combines neural models with symbolic reasoning to improve correctness and interpretability~\cite{desmet2025defining}. Examples range from logic-programming frameworks such as DeepProbLog~\cite{manhaeve2018deepproblog} to systems translating natural language into formal representations processed by symbolic solvers~\cite{liu2023llmp,pan2023logiclm,song2024leancopilot}.

Recent work explores this paradigm in system-level contexts. Approaches such as OptiGuide show that language models can generate optimisation models from high-level descriptions~\cite{li2023optiguide}, while other works integrate SMT solvers and structured reasoning pipelines. These studies consistently show that language models are effective at structuring problems but unreliable as standalone reasoners.

Neuro-symbolic approaches have also been applied to cloud and edge environments. Modular architectures combining neural and symbolic components have been proposed for general reasoning tasks~\cite{vanbekkum2021modular}, while constraint-based approaches combining language models with SMT solvers have been explored for structured reasoning problems~\cite{hsia2026neurosymbolic}. More recent work investigates neuro-symbolic methods for intent-based service management, where language models map high-level intents to structured representations for symbolic components~\cite{colombi2025neurosymbolic}.

At the same time, deploying large language models in edge environments raises concerns related to latency, cost, and resource constraints, motivating the use of smaller models and lightweight reasoning strategies~\cite{semerikov2025edge}.

\subsection{Positioning of this Work}

Declarative placement offers inspectable reasoning but requires manually specified models, while neuro-symbolic systems often address general reasoning rather than placement-specific constraints. This work connects the two by exposing Prolog as a \textit{skill}: the language model structures placement intent, and Prolog validates representations, solves constraints, and explains outcomes through a shared symbolic layer.

\section{Proposed Methodology}
\label{sec:methodology}

This work proposes a \textit{Prolog skill} for constraint-aware service placement. The skill separates intent grounding from reasoning: the language model populates a schema-constrained knowledge base, while Prolog validates, queries, and explains it.
The envisioned workflow consists of three stages:
\begin{enumerate}
	\item \textit{Schema mapping}. The meta-schema declares 
       \begin{code}
schema(node, [nodeId]).
schema(service, [serviceId]).
schema(capacity, [nodeId, resource, integer]).
schema(requires, [serviceId, resource, integer]).
schema(latency, [nodeId, nodeId, integer]).
schema(place, [serviceId, nodeId]).

generated(node(edge1)).
generated(capacity(edge1, cpu, 8)).
generated(service(videoAnalytics)).
generated(requires(videoAnalytics, cpu, 4)).
\end{code}

    the predicates that the language model may instantiate for infrastructure, services, and requested placements, separating the skill interface from request-specific facts.

	\item \textit{Guardrail validation}. Guardrails are Prolog rules, so syntax and semantic consistency are checked inside the same symbolic model.
	      \begin{codenum}[firstnumber=1]
validFact(Fact) :-
  Fact =.. [Predicate | Arguments],
  schema(Predicate, Types),
  same_length(Arguments, Types),
  wellTyped(Arguments, Types).

violation(unknownNode(Node)) :-
  generated(capacity(Node, Resource, Capacity)),
  \plnot generated(node(Node)).

violation(lowCapacity(Service, Node, Resource)) :-
  generated(place(Service, Node)),
  generated(requires(Service, Resource, Required)),
  generated(capacity(Node, Resource, Available)),
  Required > Available.

safeKnowledgeBase :-
  \plnot (generated(Fact), \plnot validFact(Fact)),
  \plnot violation(Reason).
\end{codenum}

	\item \textit{Placement reasoning}. After validation, placement queries reuse the same facts and expose violated predicates as explanations.
	      \begin{codenum}
admissibleHost(Service, Node) :-
  safeKnowledgeBase,
  generated(service(Service)),
  generated(node(Node)),
  \plnot (generated(requires(Service, Resource, Required)),
      \plnot hasCapacity(Node, Resource, Required)).

hasCapacity(Node, Resource, Required) :-
  generated(capacity(Node, Resource, Available)),
  Available >= Required.

explainFailure(Service, Node, Reason) :-
  generated(place(Service, Node)),
  violation(Reason).
\end{codenum}

\end{enumerate}

The meta-schema can also evolve under Prolog control: proposed predicates are accepted only with a valid signature and guardrail. Overall, the skill exposes a compact symbolic layer where validation, reasoning, schema evolution, and explanation share the same representation.

\section{Research Questions and Evaluation Plan}
\label{sec:evaluation}

This work is currently at an early stage and focuses on assessing the feasibility and usefulness of a Prolog-centric neuro-symbolic approach to service placement.
The evaluation is guided by four research questions, each directly associated with a specific assessment strategy.

\begin{enumerate}[label=\textit{RQ\arabic{*})}]
	\item \textit{(Expressiveness)} Which classes of placement constraints can the Prolog schema capture without extension? Assessed by its ability to represent infrastructure properties, service requirements, and constraint types across representative deployments.

	\item \textit{(Faithfulness)} How faithfully can a language model map placement intent to symbolic representations? Assessed by matching generated and reference knowledge bases to detect missing or inconsistent constraints.

	\item \textit{(Explainability)} Which satisfied and violated constraints can be traced to explicit predicates? Assessed by analysing Prolog reasoning traces and their ability to expose satisfied and violated constraints.

	\item \textit{(Orchestration utility)} Under which orchestration conditions are the resulting placements usable as decision support? Assessed by evaluating their usability as decision support in realistic deployment scenarios.
\end{enumerate}

\noindent A prototype will be developed consisting of a Prolog-based placement knowledge base and a language-model interface for intent grounding. The goal is to assess whether a Prolog skill can act as an effective intermediate layer between high-level placement intent and constraint-based deployment reasoning.

\section{Discussion and Outlook}
\label{sec:outlook}

This work proposes a Prolog-centric neuro-symbolic approach for service placement in the compute continuum. Its core contribution is a schema-constrained Prolog skill that maps placement intent to validated symbolic facts and queries. The proposed approach exposes Prolog as a skill that can be invoked by a language model to perform structured reasoning tasks for service placement.

Using Prolog as a unified layer for validation, reasoning, and explanation enables a consistent treatment of placement constraints. A shared symbolic representation is used throughout the workflow, reducing semantic mismatches and improving transparency, as decisions can be traced back to explicit facts and rules. Symbolic guardrails further support validation by encoding domain invariants within the reasoning model, although their effectiveness depends on constraint completeness and may introduce search overhead at scale.

A key aspect of the design is the separation between intent grounding and reasoning. The language model does not perform placement reasoning directly, but orchestrates the Prolog skill by generating structured symbolic representations and queries. Constraint evaluation is fully delegated to Prolog. This improves reliability compared to end-to-end neural approaches, but introduces a dependency on the correctness of the generated knowledge base, where missing or incorrect constraints may lead to unintended outcomes unless rejected by symbolic guardrails.

The approach focuses on feasibility and explanation rather than optimisation, which supports interpretability but may limit applicability in large-scale scenarios without optimisation backends or controlled schema extensions. Overall, the design highlights a trade-off between expressiveness, interpretability, and scalability, emphasising skill-based interaction between language models and symbolic reasoning as an alternative to fully neural decision-making.

Future work will focus on improving translation robustness, extending symbolic validation mechanisms, integrating optimisation capabilities, and evaluating the approach on larger and more realistic deployment scenarios.

\section*{Acknowledgment}
\noindent This work has been partially funded by the NOUS (A catalyst for EuropeaN ClOUd Services in the era of data spaces, high-performance and edge computing) HORIZON-CL4-2023-DATA-01-02 project, G.A. n. 101135927.

\AtNextBibliography{\footnotesize}
\printbibliography

@inproceedings{bisicchia2022distributed,
	author = {Bisicchia, Giuseppe and Forti, Stefano and Colla, Alberto and Pisa, Claudio and Barchiesi, Alessandro and Brogi, Antonio},
	title = {A Distributed Tool for Monitoring and Benchmarking a National Federated Cloud},
	year = {2024},
	journal = {Communications in Computer and Information Science},
	volume = {1845 CCIS},
	pages = {92 – 112},
	doi = {10.1007/978-3-031-68165-3_5},
}

@inproceedings{bisicchia2023customisable,
  author       = {Giuseppe Bisicchia and
                  Stefano Forti and
                  Alberto Colla and
                  Antonio Brogi},
  title        = {Customisable Fault and Performance Monitoring Across Multiple Clouds},
  booktitle    = {Proceedings of the 13th International Conference on Cloud Computing
                  and Services Science, {CLOSER} 2023},
  pages        = {212--219},
  year         = {2023},
  doi          = {10.5220/0011849500003488}
}

@article{shi2016edge,
	title        = {{Edge Computing: Vision and Challenges}},
	author       = {Weisong Shi and Jie Cao and Quan Zhang and Youhuizi Li and Lanyu Xu},
	year         = 2016,
	journal      = {IEEE IoT-J},
	volume       = 3,
	number       = 5,
	doi          = {10.1109/JIOT.2016.2579198}
}

@article{satyanarayanan2017emergence,
	title        = {{The Emergence of Edge Computing}},
	author       = {Mahadev Satyanarayanan},
	year         = 2017,
	journal      = {Computer},
	volume       = 50,
	number       = 1,
	doi          = {10.1109/MC.2017.9}
}

@inbook{mahmud2018fog,
	title        = {{Fog Computing: A Taxonomy, Survey and Future Directions}},
	author       = {Mahmud, Redowan and Kotagiri, Ramamohanarao and Buyya, Rajkumar},
	year         = 2018,
	booktitle    = {IoE: Algorithms, Methodologies, Technologies and Perspectives},
	publisher    = {Springer},
	doi          = {10.1007/978-981-10-5861-5_5}
}

@misc{liu2023llmp,
	title        = {{LLM+P: Empowering Large Language Models with Optimal Planning Proficiency}},
	author       = {Bo Liu and Yuqian Jiang and Xiaohan Zhang and Qiang Liu and Shiqi Zhang and Joydeep Biswas and Peter Stone},
	year         = 2023,
	eprint       = {2304.11477},
	archiveprefix = {arXiv},
	primaryclass = {cs.AI}
}

@misc{pan2023logiclm,
	title        = {{Logic-LM: Empowering Large Language Models with Symbolic Solvers for Faithful Logical Reasoning}},
	author       = {Liangming Pan and Alon Albalak and Xinyi Wang and William Yang Wang},
	year         = 2023,
	eprint       = {2305.12295},
	archiveprefix = {arXiv},
	primaryclass = {cs.CL}
}

@misc{vanbekkum2021modular,
	title        = {{Modular Design Patterns for Hybrid Learning and Reasoning Systems: a taxonomy, patterns and use cases}},
	author       = {Michael van Bekkum and Maaike de Boer and Frank van Harmelen and André Meyer-Vitali and Annette ten Teije},
	year         = 2021,
	eprint       = {2102.11965},
	archiveprefix = {arXiv},
	primaryclass = {cs.AI}
}

@misc{manhaeve2018deepproblog,
	title        = {{DeepProbLog: Neural Probabilistic Logic Programming}},
	author       = {Robin Manhaeve and Sebastijan Dumancic and Angelika Kimmig and Thomas Demeester and Luc De Raedt},
	year         = 2018,
	eprint       = {1805.10872},
	archiveprefix = {arXiv},
	primaryclass = {cs.AI}
}

@misc{song2024leancopilot,
	title        = {{Lean Copilot: Large Language Models as Copilots for Theorem Proving in Lean}},
	author       = {Peiyang Song and Kaiyu Yang and Anima Anandkumar},
	year         = 2025,
	eprint       = {2404.12534},
	archiveprefix = {arXiv},
	primaryclass = {cs.AI}
}

@article{lera2024availability,
	title        = {{Availability-Aware Service Placement Policy in Fog Computing Based on Graph Partitions}},
	author       = {Lera, Isaac and Guerrero, Carlos and Juiz, Carlos},
	year         = 2019,
	journal      = {IEEE IoT-J},
	volume       = 6,
	number       = 2,
	doi          = {10.1109/JIOT.2018.2889511}
}

@inproceedings{colombi2025neurosymbolic,
	title        = {{Investigating Neurosymbolic AI for Intent-based Service Management}},
	author       = {Colombi, Lorenzo and Cavicchi, Sara and Poltronieri, Filippo and Tortonesi, Mauro and Stefanelli, Cesare and Varga, Pal},
	year         = 2025,
	booktitle    = {CNSM},
	doi          = {10.23919/CNSM67658.2025.11297547}
}

@misc{desmet2025defining,
	title        = {{Defining neurosymbolic AI}},
	author       = {Lennert De Smet and Luc De Raedt},
	year         = 2025,
	eprint       = {2507.11127},
	archiveprefix = {arXiv},
	primaryclass = {cs.AI}
}

@article{semerikov2025edge,
	title        = {{Edge Intelligence Unleashed: A Survey on Deploying Large Language Models in Resource-Constrained Environments}},
	author       = {S. O. Semerikov and others},
	year         = 2025,
	journal      = {JEC},
	volume       = 4,
	number       = 2,
	doi          = {10.55056/jec.1000}
}

@inproceedings{massa2022dataaware,
	title        = {{Data-Aware Service Placement in the Cloud-IoT Continuum}},
	author       = {Massa, Jacopo and Forti, Stefano and Brogi, Antonio},
	year         = 2022,
	booktitle    = {Service-Oriented Computing},
	publisher    = {Springer},
	doi          = {10.1007/978-3-031-18304-1_8}
}

@article{massa2025edgewise,
	title        = {{Combining declarative and linear programming for application management in the cloud-edge continuum}},
	author       = {Jacopo Massa and Stefano Forti and Patrizio Dazzi and Antonio Brogi},
	year         = 2026,
	journal      = {FGCS},
	volume       = 176,
	doi          = {10.1016/j.future.2025.108224}
}

@article{brogi2020fogtorch,
	title        = {{QoS-aware Deployment of IoT Applications Through the Fog}},
	author       = {A. Brogi and S. Forti},
	year         = 2017,
	month        = 10,
	journal      = {IEEE IoT-J},
	volume       = 4,
	number       = 5,
	doi          = {10.1109/JIOT.2017.2701408}
}

@misc{li2023optiguide,
	title        = {{Large Language Models for Supply Chain Optimization}},
	author       = {Beibin Li and Konstantina Mellou and Bo Zhang and Jeevan Pathuri and Ishai Menache},
	year         = 2023,
	eprint       = {2307.03875},
	archiveprefix = {arXiv},
	primaryclass = {cs.AI}
}

@misc{hsia2026neurosymbolic,
	title        = {{Neuro-Symbolic Compliance: Integrating LLMs and SMT Solvers for Automated Financial Legal Analysis}},
	author       = {Yung-Shen Hsia and Fang Yu and Jie-Hong Roland Jiang},
	year         = 2026,
	eprint       = {2601.06181},
	archiveprefix = {arXiv},
	primaryclass = {cs.AI}
}

@article{forti2022declarative,
  title={Declarative continuous reasoning in the cloud-IoT continuum},
  author={Forti, Stefano and Bisicchia, Giuseppe and Brogi, Antonio},
  journal={Journal of Logic and Computation},
  volume={32},
  number={2},
  pages={206--232},
  year={2022},
  publisher={Oxford University Press}
}

@article{bisicchia2024continuous,
  title={Continuous QoS-compliant orchestration in the Cloud-Edge continuum},
  author={Bisicchia, Giuseppe and Forti, Stefano and Pimentel, Ernesto and Brogi, Antonio},
  journal={SPE},
  volume={54},
  number={11},
  pages={2191--2213},
  year={2024},
  publisher={Wiley},
  doi={10.1002/spe.3334}
}

\end{document}